\documentclass[12pt]{article}   
\usepackage{amsfonts}      
\def\debut{ \begin{eqnarray} }
\def\fin{ \end{eqnarray} }
\def\non{ \nonumber }
\def\d{\partial}      
\def\str{{\rm STr \;}}      
\def\({\left(}      
\def\){\right)}
\def\dx{ \frac{d^2 x}{2\pi} }
\def\vep{\varepsilon}
\def\psib{\bar \psi}
\def\zbar{{\bar z}}
\def\inv#1{{1 \over #1}}
\def\betab{\bar{\beta}}

\def\vev#1{\langle #1 \rangle}
\def\al{\alpha}

\def\vep{\varepsilon}
\def\la{\lambda}        
\def\de{\delta}         
         
        \def\Sig{\Sigma}

              \def\CC{{\cal C}}
\def\CD{{\cal D}}       \def\CE{{\cal E}}

              \def\CO{{\cal O}}
              
\def\CS{{\cal S}}       \def\CT{{\cal T}}

\begin{document}      
      
\centerline{\LARGE Large N Spin Quantum Hall Effect.}       
\vskip 1cm      
\centerline{D. Bernard\footnote{Member of the C.N.R.S.}, 
N. Regnault and D. Serban} 
\medskip
\centerline{\it 
Service de Physique Th\'eorique de Saclay\footnote{\it 
Laboratoire de la Direction des Sciences de la
Mati\`ere du Commisariat \`a l'Energie Atomique, URA 2306 du C.N.R.S..} 
F-91191, Gif-sur-Yvette, France.   }  
\vskip2cm

\begin{abstract}
We introduce a large $N$ version of the spin quantum Hall transition
problem. It is formulated as a problem of Dirac fermions coupled 
to disorder, whose Hamiltonian belong to the symmetry class $C$. 
The fermions carry spin degrees of freedom valued in the
algebra $sp(2N)$, the spin quantum Hall
effect corresponding to $N=1$.  Arguments based on renormalization group 
transformations as well as on a sigma model formulation, valid in
the large $N$ limit, indicate the existence of a crossover as
$N$ varies. Contrary to the $N=1$ case, the large $N$ models
are shown to lead to localized states at zero energy. We also
present a sigma model analysis for the system of Dirac fermions
coupled to only $sp(2N)$ random gauge potentials, 
which reproduces known exact results.
\end{abstract}
\vfill
\newpage

\section{Introduction.}

The landscape of delocalization transitions is wider 
in two dimensions than in higher dimensions, 
paralleling the classification of the new random 
ensembles of ref.\cite{AZ}.
Many of these transitions may be modeled by Dirac fermions
coupled to various disordered variables. 
When formulated as field theories these transitions are 
usually mapped into difficult strong coupled systems requiring
the identification of non-trivial infrared fixed points.
In this respect, the recently introduced $su(2)$ spin quantum
Hall transition \cite{sqhe}, which corresponds to random 
Hamiltonians belonging to class C of ref.\cite{AZ}, appears
as an exception. Indeed, using a network formulation,
it has been argued \cite{GLR} that the critical properties
of the latter model are potentially described by percolation in 
two dimensions. This opened the possibility of exact computations
of characteristic critical exponents \cite{GLR} or of the mean
conductance \cite{cardy}.
See also ref.\cite{SMF} for a spin chain formulation of this model.
However, these results still resist to a field theory derivation,
see eg. \cite{two}. Such alternative approach would be useful
for the related, but yet unsolved, problem of the quantum
Hall transition. 

The aim of this paper is to study a large $N$ version of the
spin quantum Hall transition. We shall formulate it as
Dirac fermions with $sp(2N)$ spin degrees of freedom 
whose Hamiltonians belong to the class $C$. This requirement,
which forces us to choose the algebra $sp(2N)$, is compatible with
the introduction of four types of disordered potentials. These potentials
are all generated under renormalization group transformations. 
Some of the results valid in the $N=1$ spin quantum Hall effect
extend to the large $N$ generalization. This holds true for
instance for the spin-charge separation for particular
fine tuned versions of the models. However the following
analysis indicates that there exists a crossover as $N$ is increased.
Namely, we will find that, contrary to the $N=1$ case, 
zero energy states are localized at large $N$.

We analyse these models in two steps. First, renormalization 
group transformations allow us to identify the field theory describing 
the universality class of such models. It corresponds to 
disordered variables isotropically distributed among the
four possible types of disordered potentials. We then use
supersymmetric sigma model techniques to analyse the 
universal model, which for class $C$ Dirac fermions
is a sigma model on the Riemannian symmetric superspace $OSp(2|2)/GL(1|1)$, 
of type $D{\rm III}|C{\rm I}$ according to ref.\cite{riemann}.
This sigma model, characteristic for the symmetry class
$C$, also appeared in the context of disordered superconductors
with spin rotation symmetry and no time reversal
invariance \cite{vortex,asz}.
The resulting sigma model turns out to be a massive theory.
As a consequence the low
energy limit is non critical. This is 
in contrast with class $D$ Dirac fermions which were
described in refs.
\cite{supraD,classD} by a massless sigma model 
of type $C{\rm I}|D{\rm III}$.
In both cases, the derivation of the sigma model is
justified only at large $N$.

For completeness, we have also included a sigma model description
of Dirac fermions coupled to $sp(2N)$ random gauge potentials. This
corresponds to a particular fine tuning of the previously
discussed models  enlarging the symmetry of the Hamiltonians
which then belongs to class $C{\rm I}$. Surprisingly, large $N$ sigma 
model analysis reproduces the exact result \cite{two} that
the critical theory is an $osp(2|2)_{k=-2N}$ WZW model.
The fixed point of class CI, including the WZW term, was also
found in \cite{NTW,Fend} using replica and in \cite{beber} by direct means.

The paper is organized as follows.
The models as well as their supersymmetric formulations are introduced
in Section 2 and 3. There, we also introduce the appropriate
orthosymplectic transformations and the effective actions. Symmetries
of the pure systems are identified in Section 4. The beta functions
computed in Section 5 show that the model with isotropic randomness 
is universal and attractive at large $N$. The sigma model formulations 
of the latter, which is valid in the large $N$ limit, is analysed in
Section 6. We show that, at large scale, this model is driven to
a strongly coupled massive phase, due to the absence of 
non-trivial topological $\theta$-term. This leads to our main result
concerning localization of zero energy eigenstates. Spin-charge
separation in the pure system and sigma model formalism for perturbations
by random gauge potentials, corresponding to random Hamiltonians in
class $C{\rm I}$, are presented in Section 7 and 8.
\bigskip
 
{\bf Acknowledgments:} It is a pleasure to thank M. Zirnbauer
for exchanges on related matter, for pointing to us that spin-charge
separation should  extend to any $N$ and
for furnishing us ref.\cite{Howe}, and M. Bauer and A. LeClair
for numerous discussions and related collaborations.

\section{The models.}

We consider $2N$ species of Dirac fermions, coupled by disorder
in class C \cite{AZ}.  The  Dirac Hamiltonian for generic disorder 
in this class is
        \begin{equation}
        \label{hamc}
        H =  \pmatrix{\alpha+M+m & -2i \partial + A \cr
        -2i \bar \partial + \bar A & \alpha -M-m}  
        \end{equation}     
where $\d =( \d_x - i \d_y)/2 $, $\bar \d = (\d_x + i\d_y)/2 $, 
$A =  A_x - i A_y $,   $\bar A = A_x + i A_y $.
Here, $A_\mu = \sum_a A_\mu^a (x,y) \tau^a $ are random $sp(2N)$ gauge 
potentials, $\alpha=\sum_a \alpha_a (x,y) \tau^a $ is a random ``spin"
potential, and $M= \sum_i M_i (x,y) \CT^i$ and $m=m(x,y) {\bf 1} $
are random mass terms. We consider generators of $sp(2N)$ 
defined by the relation 
        $$\tau^a =-\Sigma\; (\tau^a)^T\; \Sigma^{-1}\;,
        $$
where the superscript $T$ denotes usual transposition, and $\Sigma$ is the
symplectic unit 
        \begin{equation}
        \label{msigma}
        \Sigma=\left( \matrix{0 & {\bf 1}_N \cr -{\bf 1}_N &0} 
        \right)\;.
        \end{equation}
The generators $\CT^i$, with the property  
$\CT^i =\Sigma (\CT^i)^T \Sigma^{-1}$, form the complement of $sp(2N)$ with respect
to $sl(2N)$. The generators $\tau^a$ and $\CT^i$, 
together with the identity ${\bf 1}$,
span the algebra $gl(2N)$. Remark that
 for $N=1$ there is no generator $\CT^i$, so that $sp(2)\simeq su(2)$.

The Hamiltonian (\ref{hamc}) enjoys the following particle-hole symmetry
defining symmetry class C (spin rotation invariance 
and no time reversal symmetry) in the classification of \cite{AZ} 
        \begin{equation}
        \label{classc}
        H=-\CC\; H^T \; \CC^{-1}\;,
        \end{equation}
with $\CC=\sigma_1 \otimes \Sigma$ an antisymmetric matrix.
This relation implies that the eigenvalues occur in pairs with opposite
signs, and it relates the advanced and retarded Green 
functions
        $$G_R(E)=-\CC\; G_A(-E)^T\; \CC^{-1}\;.$$

We consider centered gaussian distributions for the four types of disorder,
with strengths $g_A$, $g_\alpha$, $g_M$ and $g_m$ (positive real numbers
for real disorder). The disorder variances are:
\debut
\vev{m(x)m(y)}=\frac{g_m}{N}\,\delta^{(2)}(x-y) \quad &,&\quad
\vev{M(x)M(y)}=\frac{g_M}{N}\,\delta^{(2)}(x-y) \non\\
\vev{\al^a(x)\al^b(y)}=\frac{g_\al}{N}\,\de^{ab}\delta^{(2)}(x-y) \quad &,&\quad
\vev{A^a(x)\bar A^b(y)}=\frac{2g_A}{N}\,\de^{ab}\delta^{(2)}(x-y) \non
\fin
The reason to consider the four types of disorder is that they are generated
by renormalization of the effective action. 
As we will see in the following sections, there are particular cases
where the symmetry of the Hamiltonian (\ref{hamc}) is greater than
(\ref{classc}) and the renormalization group flow closes on a 
subset of the disorder coupling constants;
this is the case for example for the 
random vector potential alone, when the symmetry class changes to $C{\rm I}$..

The single particle Green functions are defined by the functional integral
$Z^{-1}\,\int D\Psi^* D\Psi ~ \exp (-S)$ with $Z$ the partition function and
        \begin{equation}
        \label{2.5} 
        S = \int \dx ~ \Psi^{\star} (x) i \( H - \CE \) \Psi (x) 
        \end{equation}
where $\CE = E + i\vep$.  For $\vep = 0^+$, this defines the retarded 
Green function
        \begin{equation}
        \label{2.6}
        G_R (x,x';E) = \lim_{\vep \to 0^+} \langle x | \inv{H-(E+i\vep)} |x' \rangle 
        = \lim_{\vep \to 0^+} i \langle \Psi (x) \Psi^{\star} (x') \rangle 
        \end{equation}
Letting 
        \begin{equation}
        \Psi = \left( \matrix{ \psi_+ \cr \psib_+ \cr} \right), 
         ~  \Psi^{\star} = (\psib_- , \psi_- ) \non
        \end{equation}
where $\psi_+$ is a 2N-component fermion $\psi_+^i, i=1,\ldots,2N$, 
and similarly for $\psib_\pm$, one finds 
        \begin{eqnarray}
        \nonumber
        S&=&\int \dx \Big( \psi_- 
        (2 \bar \d + i \bar  A ) \psi_+  + \psib_- (2 \d + i A ) \psib_+ 
        + i(\psi_- \alpha \psib_+ + \psib_- \alpha \psi_+ ) \nonumber
        \\  
        &+& im (\psi_- \psib_+ - \psib_- \psi_+ ) 
        + i(\psi_- M \psib_+ - \psib_-M \psi_+ )
        -i \CE \Phi_E   \Big) 
        \label{2.8}
        \end{eqnarray} 
where $\Phi_E = \psi_- \psib_+ + \psib_- \psi_+ $.  

\section{Supersymmetric effective action.}

Since the disorder is gaussian and the deterministic part of the
hamiltonian is free, we can use the supersymmetric method to
compute disorder averages of Green functions. For each fermion field,
we introduce
bosonic partners, $\beta^i_\pm, \betab_\pm^i $, 
$i=1,\ldots,2N$, so that the inverse
of the partition function for fermions become a partition function 
for bosons
        \begin{equation}
        \label{3.1}
        Z(\alpha, m ,M, A)^{-1}  = \int D\beta ~ e^{-S(\psi \to \beta)} 
        \end{equation}
In order to simplify the notation and to take advantage of
the symmetries of the problem, we are going to introduce
supermultiplets, each containing $4N$ fermion fields and $4N$ boson
fields
        $$
        \phi = \pmatrix{\psi_+ \cr \Sigma\;\psi_-^T\cr  \beta_+
         \cr \Sigma\;\beta_-^T \cr} \;, \qquad
        \bar\phi = \pmatrix{\bar\psi_+\cr\Sigma\; \bar\psi_-^T\cr 
        \bar\beta_+  \cr \Sigma \; \bar\beta_-^T \cr}\;,
        $$
with $\Sigma$ defined in eq. (\ref{msigma}).
Here the index $i=1,\ldots,2N$ was omitted, and the superscript
$T$ represents the usual transposition. $\psi_+$ is a column vector
and $\psi_-$ a row vector.
We define also orthosymplectic transposes $\phi^t$ of $\phi$
and $\bar \phi^t$ of $\bar \phi$ by
        \begin{equation}
         \phi^{\rm t} \equiv ( \psi_- , \psi_+^T\;\Sigma^{-1} ,  
        \beta_-  , -\beta_+^T\;\Sigma^{-1}  ) \;,
        \quad  \bar \phi^{\rm t} \equiv ( \bar\psi_- , \bar\psi_+^T\;\Sigma^{-1} ,  
        \bar\beta_-  , -\bar\beta_+^T\;\Sigma^{-1}  ) \;.
        \label{transpose}
        \end{equation}
The inner product associated to this transposition 
is skew
        $$\bar\phi^t \phi=-\phi^t\bar \phi\;.$$
Similarly,
\debut
 \bar \phi^t\tau^a\phi= \phi^t\tau^a\bar \phi \quad ;\quad
\bar \phi^t\CT^i\phi= -\phi^t\CT^i\bar \phi\;. \label{antisym}
\fin
One can define a transpose of $8N \times 8N$ supermatrices $A$
by $(A\phi)^t\equiv \phi^t A^t$. Its explicit relation to the usual
supertranspose (we use same symbol $T$ to denote the transposition for
the usual matrices and the supertransposition for the supermatrices), is given by
        \begin{equation}
        \label{orthotransp}
        A^t=\gamma A^T \gamma^{-1}\;, \qquad 
        \gamma=\pmatrix{0 & \Sigma^{-1}\cr \Sigma & 0} \otimes E_{FF}-
         \pmatrix{0 & \Sigma^{-1}\cr -\Sigma & 0} \otimes E_{BB}\;,
        \end{equation} 
$E_{FF}$ and $E_{BB}$ being the projectors on the fermion-fermion space and
boson-boson space respectively.
The present definition of the supermultiplets $\phi$ and $\phi^t$, as
well as of the orthosymplectic transpose is different from the one
used in \cite{classD}, and it is adapted to the present type of disorder.
To be able to compare the symmetry properties of supervectors
and supermatrices in the two cases it useful to retain the following 
relations
        \begin{equation}
        \phi_{{\rm class\; C}}= \pmatrix{1& 0\cr 0&\Sigma}\otimes
{\bf 1}_{\rm susy}\,
        \phi_{{\rm class\; D}}\;,
        \qquad \phi^t_{{\rm class\; C}}=\phi^t_{{\rm class\; D}}
        \pmatrix{1& 0\cr 0&\Sigma^{-1}}\otimes {\bf 1}_{\rm susy}\;,
        \end{equation}
and similarly for $\bar \phi$ and $\bar \phi^t$.

The free theory of fermions plus the bosonic ghosts is conformal, with Virasoro 
central charge $c=0$ and with action
        \begin{eqnarray}
        \label{3.3} 
        S_{{\rm cft}} &=&2 \int \dx \( \psi_- \d_\zbar \psi_+ + \psib_- \d_z \psib_+ 
        + \beta_- \d_\zbar \beta_+ + \betab_- \d_z \betab_+ \) \\ \nonumber
        &\equiv &
        \int \dx \(\phi^t \; \d_\zbar \phi + \bar \phi^t \; \d_z \bar \phi\) \;. 
        \end{eqnarray}
The short distance singularity of the holomorphic fields in the 
supermultiplet $\phi$ can be written as
\debut
\phi(z) \phi^t (w) \simeq \frac{{\bf 1}}{z-w} \label{wick}
\fin
and similarly for the supermultiplet $\bar \phi$. 
         
After performing the integrals over the disorder, we obtain the
following effective action
        \begin{equation}
        \label{3.2} 
        S_{{\rm eff}} = S_{{\rm cft}} + 
        \frac{1}{2N} \int \dx \( g_\alpha \, \CO_\al  + g_m \, 
        \CO_m + g_M \, \CO_M+ g_A \, \CO_A
        \)\;, 
        \end{equation}
with the following operators perturbing away from the conformal field theory
        \debut
\label{oppert}
        &\CO_\al& = (\bar \phi^t\, \tau^a\, \phi )(\bar \phi^t\, \tau^a\, \phi )\;,
        \\ \nonumber
        &\CO_m& =\frac{1}{2N} (\bar \phi^t\, \phi )(\bar \phi^t \, \phi )\;,\\ 
        \noalign{\vskip2pt}
        &\CO_M& = (\bar \phi^t\, \CT^i\, \phi )(\bar \phi^t\, \CT^i\, \phi )\;,
         \nonumber \\ \noalign{\vskip5pt}
        &\CO_A& = ( \phi^t\, \tau^a\, \phi )(\bar \phi^t\, \tau^a\, \bar \phi )\;.
         \nonumber
        \fin
It will turn out to be useful to know alternative ways of
writing these operators. Using the antisymmetry properties (\ref{antisym}) 
as well as the cyclicity of the supertrace, one has:
 \begin{eqnarray}
\label{oppertbis}
        &\CO_\al& = 
{\rm STr}(\,\phi\bar \phi^t\,\tau^a\,\phi\bar \phi^t\, \tau^a\,)=
{\rm STr}(\,\bar \phi\bar \phi^t\,\tau^a\,\phi \phi^t\, \tau^a\,)
        \\ \nonumber
        &\CO_m& =
\frac{1}{2N}{\rm STr} (\, \phi\bar \phi^t\, \phi\bar \phi^t\,)=
-\frac{1}{2N}{\rm STr} (\, \bar \phi\bar \phi^t\, \phi \phi^t\,) 
       \\   \noalign{\vskip2pt}
        &\CO_M& = 
{\rm STr}(\,\phi\bar \phi^t\, \CT^i\, \phi \bar \phi^t\, \CT^i\,)=
-{\rm STr}(\,\bar \phi\bar \phi^t\, \CT^i\, \phi \phi^t\, \CT^i\,)
         \nonumber \\ \noalign{\vskip5pt}
        &\CO_A& = 
{\rm STr}(\, \bar \phi \phi^t\, \tau^a\, \phi \bar \phi^t\, \tau^a\,)\;.
         \nonumber
        \end{eqnarray}

Finally, one can consider the energy term, which can be written
in the present notations as
        \begin{equation}
        \Phi_E=(\bar \phi^t\, \Sigma_3\, \phi)\;,
        \end{equation}
where $\Sigma_3={\bf 1}_{2N}\otimes \sigma_3 \otimes {\bf 1}_{{\rm susy}}$.
In order for the integrals over the bosonic field to converge,
one has to relate the bosonic fields to each other by complex conjugation
\debut
\beta_-^{\dagger}=\bar \beta_+\;, \qquad \beta_+^{\dagger}=\bar
\beta_-\;.
\label{real}
\fin
Then, a positive imaginary part of the energy insures convergence
of the bosonic functional integrals.

\section{Symmetries.}

Before perturbation by the disorder operators (\ref{oppert}), 
the action (\ref{3.3}) possesses conformal invariance
 with current algebra symmetry \cite{KZ}.
Bilinears in the chiral supermultiplets $(\phi\, \phi^t)$
generate an $osp(4N|4N)$ algebra. 

Two natural subalgebras of $osp(4N|4N)$ are $sp(2N)$ and $osp(2|2)$.
The $sp(2N)$ algebra, which we refer to as the spin algebra, 
is generated by the currents,
 \begin{equation} 
 J^a(z) = (\phi^t\, \tau^a\, \phi)(z) \label{spcurrent}
 \end{equation}
They satisfy
\debut
J^a(z)\, J^b(0) \simeq \frac{f^{ab}_c}{z}J^c(0) + \ {\rm reg.}
\non
\fin
with $f^{ab}_c$ the $sp(2N)$ structure constants,
showing that they form a representation of the affine algebra $sp(2N)$ 
at level zero. 

The $osp(2|2)$ algebra, which we refer to as the charge algebra, 
is generated by the $sp(2N)$ scalar
\debut
K(z)= {\rm Tr}_{gl(2N)}(\phi\,\phi^t)_{(z)} 
={\rm Tr}_{gl(2N)}(E_I\,(\phi\,\phi^t)_{(z)}\, E_I) \label{ospcurrent}
\fin
where the trace $Tr_{gl(2N)}$ is over the spin indices and
$E_I$ is any orthonormalized basis of $gl(2N)$. 
A convenient basis is be made of $\tau^a,\ \CT^i$ and 
the identity ${\bf 1}$. The currents
$K$ may be viewed as a $4\times4$ supermatrix, 
$K_{\al\beta}=\sum_i \phi_{i\al}\phi^t_{i\beta}$, with components:
\debut
J = \psi_+^i\psi^i_- \quad &,& \quad
J_\pm = \Sig_{ij} \psi_\pm^i\psi_\pm^j \label{ospope}\\
H=\beta_+^i\beta_-^i \quad &,& \quad 
S_\pm= \psi_\pm^i\beta_\mp^i \quad, \quad 
\hat S_\pm = \Sig_{ij} \beta_\pm^i\beta_\pm^j \non
\fin 
The even part of $K$ is made of two blocks $K_{FF}$ and $K_{BB}$
with:
$$
        K_{FF}=\pmatrix{ J & J_+ \cr J_- & -J \cr}
        \quad ,\quad
        K_{BB} = \pmatrix{ H & 0\cr 0 & -H \cr}\;.
$$ 
The currents (\ref{ospope}) generate an $osp(2|2)_k$ current algebra at level
$k=-2N$, cf ref.\cite{two}.
In particular, the $K_{BB}$ block generates a $so(2)$ algebra and 
the $K_{FF}$ block generates a $sp(2)$ algebra.

The $osp(2|2)$ generators commute with the $sp(2N)$ currents:
\debut
 [ J^a\, , \, K ] =0\;. \non
\fin

The supermultiplets $\phi$ and $\phi^t$ transform
as affine  primary fields with value in the tensor product of
the defining vector representation of $sp(2N)$ by the $4$-dimensional 
representation of $osp(2|2)$. One may alternatively \cite{Howe}
view the field $\phi$ as a rectangular supermatrix with components $\phi_{i\al}$,
$i=1,\cdots,2N$, $\al=1,\cdots,4$, on which the algebra $sp(2N)$ acts
by multiplication on the left, while $osp(2|2)$ acts by multiplication
on the right. These actions of course commute.

\section{One loop beta functions.}

The first step in studying the effect of the disorder perturbation in
(\ref{3.2}) is to analyse the renormalisation group flow 
for the coupling constants.
Even at one loop, beta functions code for important properties of the system.
They  can be deduced from the operator product 
expansions (OPE) of marginal perturbing operators ${\cal{O}}_i$, 
where $i$ stands for one of the indices of 
the disorder operators (\ref{oppert}). Using the following notation for OPE
\debut 
{\cal{O}}_i(z,\bar{z}) {\cal{O}}_j(0)
&\simeq&\frac{1}{z\bar{z}} C^k_{ij} {\cal{O}}_k(0)+ {\rm reg.}\non
\fin
the one loop beta functions are given by \cite{Zam,Car}
\debut
\beta_k\equiv l\d_l\, g_k &=&-\frac{1}{2N}\sum_{i,j} C^k_{ij} g_i g_j\non
\fin
After some careful calculations, we obtain the following results
\debut
\beta_m &=& \frac{(2N+1)}{N}\Big({2g_A(g_m+g_\alpha)-g_mg_\alpha 
-\frac{(N-1)}{N}g_m g_M }\Big)-\frac{1}{N^2}g_m^2\non\\
\beta_A &=& \frac{1}{2} (g_M + g_\alpha)^2
+ \frac{1}{2N}(g_\alpha^2 - g_M^2) 
+ \frac{2(N+1)}{N}g_A^2 + \frac{1}{N^2}g_\alpha(g_m - g_M)\non\\
\beta_\alpha &=& 2g_A(g_\alpha + g_M) + \frac{1}{N}g_\alpha(g_M - g_\alpha)
 +\frac{1}{N^2}\Big({g_\alpha + 2g_A }\Big)(g_m-g_M)\non\\
\beta_M &=& \frac{2(N+1)}{N}g_A(g_\alpha+g_M) + \frac{1}{N}g_M  (g_M - g_\al)
+ \frac{1}{N^2}g_M (g_M - g_m)\label{beta1loop}
\fin
Readers who are interested in details can find some hints in the Appendices.

From now on, we shall be interested in the large $N$ limit.
The beta functions (\ref{beta1loop}) then simplify dramatically
\debut
\beta_m = -2 g_m (g_\alpha+g_M) + 4 g_A (g_\alpha+g_m) &, &
\beta_A = \frac{1}{2} (g_\alpha+g_M)^2 + 2 g_A^2 \non\\
\beta_ \alpha= 2 g_A (g_\alpha+g_M) &, & \beta_M= 2 g_A (g_\alpha+g_M)
\label{lnbeta1loop}
\fin
Of course the large $N$ limit only applies for $g\ll N$.

After a large number of RG iterations, RG trajectories
are asymptotic to directions which are  preserved by the RG flow
 and which pass through the origin. One can show that there
exist six stable directions. 
But only one is attractive in the region where coupling constants are
positive. This direction, which we call the ``isotropic direction'',
 corresponds to the case where all coupling constants are equal: 
$g_m=g_\al=g_M=g_A=g$. The associated beta function is 
\debut 
\beta_g=4g^2 \label{betaiso}
\fin 

To show the isotropic direction is attractive, we use the same method
as in \cite{two}. We project RG flow on the sphere and parameterize 
coupling constants with a radial coordinate $\rho$ and 
three angle coordinates $\theta_i$. We rewrite the RG equations as 
\debut
\dot{\theta}_i = \rho\beta_i(\theta_1, \theta_2, \theta_3) &\rm{and}&
\dot{\rho}=\rho^2\beta_\rho(\theta_1, \theta_2, \theta_3)\non
\fin
We develop the $\beta_i$ around the isotropic direction, 
and compute the corresponding eigenvalues. 
These eigenvalues are all negative, proving that  the direction is
attractive. Furthermore, it can be shown that $\dot{\rho}>0$.
These results are corroborated by the all order computations
of Appendix C. 

Thus, the isotropic direction describes the universality class 
in the region where coupling constants are all positive. 
It corresponds to a strong coupled system that we will study in details in the 
following section.

\section{The sigma model approach.}

When the disorder couplings are all positive (and $N$ is large), 
the renormalization group flow is attracted
towards the line $g_\al=g_m=g_M=g_A=g$ (``isotropic disorder'').
On this line, preserved by the flow, the low energy physics can be 
described by a non-linear sigma model with a topological term.
In this section, we derive the effective action for this
sigma model. Its target space is the Riemannian symmetric superspace 
$OSp(2|2)/GL(1|1)$, denoted $D{\rm III}|C{\rm I}$ in \cite{riemann} 
(note that there exists another $OSp(2|2)/GL(1|1)$ symmetric superspace,
denoted $C{\rm I}|D{\rm III}$, appearing in the
context of the symmetry class $D$ \cite{classD}).

When all the couplings are equal, the perturbation term
in the Lagrangian, quartic in the Dirac fields
can be decoupled by a Hubbard-Stratonovich transformation involving 
an unique supermatrix field $Q$.
Using the fact that $(\phi^t\, \phi )=
(\phi^t\, \CT^i\, \phi )=0$ and denoting
$$B=\bar \phi \phi^t+\phi \bar \phi^t$$ 
the perturbation term becomes
        \begin{eqnarray}
        L_{{\rm pert}}&=&\frac{g}{2N}\;\bigg(
         (\bar \phi^t\, \tau^a\, \phi )(\bar \phi^t\, \tau^a\, \phi )+ 
        \frac{1}{2N} (\bar \phi^t\, \phi )(\bar \phi^t \, \phi )\\ 
        \nonumber
        &+&(\bar \phi^t\, \CT^i\, \phi )(\bar \phi^t\, \CT^i\, \phi )+
        (\bar \phi^t\, \tau^a\, \bar \phi )
        ( \phi^t\, \tau^a\,  \phi ) \bigg)\\
         \nonumber
        &=&\frac{g}{4N}\; \str B \bigg(\tau^a B\; \tau^a+\frac{1}{2N} B+
        \CT^i B\; \CT^i \bigg)\;.
        \end{eqnarray}
The supermatrix $B$ obeys the relation $B=-B^t$, relation which
defines an element of the algebra $osp(4N|4N)$.
Conjugation by the generators of the subalgebra
$gl(2N)$ projects $B$ on the identity on this subalgebra
        $$\tau^a B\; \tau^a+\frac{1}{2N} B+
        \CT^i B\; \CT^i=\({\rm Tr}_{gl(2N)} B \) \otimes {\bf 1}_{2N}\;,$$
so that the perturbation term reads
        $$
        L_{{\rm pert}}=\frac{g}{4N}\;\str \({\rm Tr}_{gl(2N)}\; 
        (\bar \phi \phi^t+ \phi \bar \phi^t)\)^2\;.$$
This interaction can be decoupled using a supermatrix 
belonging to the $osp(2|2)$ algebra, $Q \sim {\rm Tr}_{gl(2N)}\; 
(\bar \phi \phi^t+ \phi \bar \phi^t)$. The resulting effective 
lagrangian is
        \begin{eqnarray} 
         L_{{\rm eff}} &=& \bar\phi^t \partial \bar\phi 
         + \phi^t \bar\partial \phi + {\rm STr} \,\( Q 
        {\rm Tr}_{gl(2N)} (\phi \bar\phi^t + \bar\phi 
        \phi^t)\) - \frac{N}{g} \, {\rm STr} \, Q^2  \nonumber \\ 
         &=& (\bar\phi^t\,\phi^t) \pmatrix{Q &\partial\cr 
        \bar\partial &Q\cr} \pmatrix{\phi\cr \bar\phi\cr} - 
        \frac{N}{g}\,{\rm STr} \, Q^2 \;. \label{decouple} 
        \end{eqnarray}  
The next step is to perform the gaussian integrals over the Dirac fields, 
resulting in the following effective action 
        \begin{equation} 
        S[Q] = - {N \over g}\int \frac{d^2x}{2\pi} \,{\rm STr}\,Q^2 + {N }\, 
        {\bf STr}\ln\pmatrix{\Sigma_3 Q &\partial\cr \bar\partial &Q\Sigma_3\cr}\;,
        \label{action} 
        \end{equation} 
where ${\bf STr}$ combines the operations of taking the supertrace 
${\rm STr}$ over the matrix indices 
and integrating over position space.  The factors 
$\Sigma_3$ under the logarithm appear after the transformation  
$\bar\phi \mapsto \Sigma_3 \bar\phi$ and $\bar\phi^{\rm t} \mapsto   
 \bar \phi^{\rm t} \Sigma_3$, correcting for the fact that  
the complex conjugate of $\bar \phi$ is not $\phi^t$ 
but $\phi^{\rm t}\Sig_3$, see eq.(\ref{real}). 
The number $N$ appears now as a factor in the action, suggesting
to treat the integral in the saddle point approximation.
The saddle point equation for the action (\ref{action}) is given by
        \begin{equation}
        \frac{2Q(x)}{g}={\rm Tr}_{gl(2)} \left\langle x \left|
         \CD^{-1} \right| x \right\rangle\;, \quad {\rm with}
        \quad \CD=\pmatrix{ Q &\partial\cr \bar\partial &Q\cr}
        \end{equation} 
We look for a spatially homogeneous solution of the form $Q(x) =
\mu \Sigma_3$.  The saddle-point equation then reduces to
         $$
         g^{-1} = \int {d^2 k \over 2\pi} \, (\mu^2 + k^2/4)^{-1} \;.
        $$
Cutting off the integral in the ultraviolet by $|k| < 1/\ell_0$ yields
the equation
\debut
        1 /2 g = \ln \left( 1 + (2\mu\ell_0)^{-2} \right)
   \approx -2 \ln (2\mu\ell_0)   \;,   
\label{gsaddle} 
\fin 
and by inversion,
        \begin{equation}
        \mu = (2\ell_0)^{-1} / \sqrt{{\rm e}^{1/2g} - 1}
\approx (2\ell_0)^{-1}\, {\rm e}^{-1/4g} \;,
        \label{genmass}
\end{equation}
As the dynamically generated mass $\mu$ is a renormalization group
invariant, eq.(\ref{genmass}) corresponds to a beta function
$\beta_g = 4g^2$ in agreement with eq.(\ref{betaiso}).
\medskip

{\it Symmetry of $Q$ and the saddle point manifold.} The bilinear in the 
Dirac fields $B=\bar \phi \phi^t+\phi \bar \phi^t$ belongs to the complex algebra
$osp(4N|4N)$ defined by the relation $B=-B^t$. The even part of this algebra 
is $sp(4N)\otimes so(4N)$, with $so(4N)$ in the fermion-fermion 
($FF$) block and 
$sp(4N)$ in the boson-boson ($BB$) block (this is due to the fact that the matrix
$\gamma$ in (\ref{orthotransp}) is symmetric in the $FF$ sector and antisymmetric 
in the $BB$ sector). 
Taking the trace over $gl(2N)$ in $B$
leaves us with an object belonging to the $osp(2|2)$ algebra, but now with the
$so(2)$ part in the $BB$ sector and the $sp(2)$ part in the $FF$ sector.
To see the way it happens, let us look at the $FF$ part in a $osp(4N|4N)$
matrix, defined by $M^t=-M$. It is easily verified that this block has the 
structure 
        \begin{equation}
        M_{FF}=\pmatrix{M_{11} & \Sigma\, M_{12} \cr M_{21}\, 
        \Sigma^{-1}& -\Sigma\, M_{11}^T\, \Sigma^{-1}}\;,
        \end{equation} 
where $M_{11}$, $M_{12}$ and $M_{21}$ are $2N\times 2N$ 
ordinary matrices obeying
$M_{12}^T=-M_{12}$ and $M_{21}^T=-M_{21}$. This implies
that $M_{FF}$ belongs to $so(4N)$. Let us now take the trace over the 
colour indices 
        \begin{equation}
        \tilde M_{FF}\equiv {\rm Tr}_{gl(2N)} M_{FF}
        =\pmatrix{\tilde M_{11} & \tilde M_{12} \cr \tilde M_{21} 
        & -\tilde M_{11}^T }\;.
        \end{equation}
Using the cyclicity of the trace and the antisymmetry of the 
matrix $\Sigma$, we can show that  $\tilde M_{12}^T=\tilde M_{12}$ and 
$\tilde M_{21}^T=\tilde M_{21}$, which means that $\tilde M_{FF}$ belongs
to the $sp(2)$ algebra. Similarly, $\tilde M_{BB}$ can be shown to belong
to $so(2)$.
The matrix $Q$ inherits this symmetry from the
object to which it couples, $\tilde B={\rm Tr}_{gl(2N)}B$, therefore it 
belongs to $osp(2|2)$.

When decoupling the interaction part with the help of the supermatrix
$Q$, one of the question that has to be addressed is the choice of the 
contour of integration. In particular, solving this question allows
to choose the acceptable solutions for the saddle
point equation, that is the ones which lie on the contour of
integration or which can be attained from it by analytical continuation. 
These questions have been addressed in detail in \cite{riemann} for the 
class $C$. There, it was shown that the dominant diagonal
saddle point is of the form 
        $$Q_0=\mu \; \Sigma_3\;,$$
with $\Sigma_3=\sigma_3\otimes {\bf 1} _{{\rm susy}}$ being
an element of $osp(2|2)$.
Due to the global $OSp(2|2)$ symmetry of the effective action, this saddle point
extends to a saddle point manifold
        $$Q_0 \to \mu\; T \Sigma_3 T^{-1}\;,$$
where $T$ is a constant element of $OSp(2|2)$. 
Since the stabilizer
of $\Sigma_3$ is $GL(1|1)$, the saddle point manifold is 
the coset space $OSp(2|2)/GL(1|1)$. This coset space can be parameterized
by $T=\exp{X}$, with $\lbrace X,\Sigma_3\rbrace=0$.

The convergence conditions for the integrals over $Q$ restrict
the saddle point manifold to a real submanifold of the
complex space $OSp(2|2)/GL(1|1)$. 
In the $FF$ sector, convergence of the integrals over $Q$ can be 
insured  by choosing $Q_{FF}^\dagger =Q_{FF}$. At the level
of the saddle point, this translates into $X_{FF}^\dagger=-X_{FF}$.
Therefore, the fermion-fermion sector of the saddle point manifold
is isomorphic to the compact symmetric
space $Sp(2)/U(1)$. The convergence conditions on $Q$ in the boson-boson sector
are more involved \cite{riemann}; on the saddle point manifold they can be reduced
to $X_{BB}^\dagger=X_{BB}$, showing that the bosonic part of the 
saddle point manifold is non-compact.
When averages of $n$ Green functions are considered, 
it is isomorphic to the non-compact symmetric space 
$SO^*(2n)/U(n)$, with $SO^*(2n)$ some real form of $SO(2n)$.
For $n=1$, the boson-boson sector is empty. 
The two symmetric spaces form the base manifold of  a Riemannian symmetric 
superspace of type $D{\rm III}|C{\rm I}$ \cite{riemann}.
\medskip

{\it Gradient expansion.}
The next step in the derivation of an effective action  
is to perform a gradient expansion of the 
action (\ref{action}). The low energy configurations are given
by the slowly varying field 
        \begin{equation}
        \label{slow}
        q(x)\equiv Q(x)/\mu =T(x)\Sigma_3 T(x)^{-1}\;,
        \end{equation}
where $T(x)$ is a (slowly varying) element of $OSp(2|2)$. Note
that $q(x)$ satisfies the nonlinear constraint $q(x)^2=1$.
The degrees of freedom $q(x)$ correspond to the  Goldstone modes
of the broken symmetry $OSp(2|2)\to GL(1|1)$.
Fluctuations transverse to the saddle point manifold  are massive 
and can be neglected at this stage.

The effective action for the Goldstone modes is a non-linear sigma
model on the symmetric superspace $OSp(2|2)/GL(1|1)$ described previously. 
This sigma model may support a 
topological term, since $\Pi_2(Sp(2)/U(1))={\bf Z}$.
The easiest way to extract the coupling constants of the kinetic and 
topological term is by using the non-abelian bosonisation \cite{witten}. 
In the supersymmetric setting, this method was used and 
explained in detail for class D in \cite{classD}
and it can be applied with minimal changes to the present case.
We want to evaluate the action (\ref{action}) on configurations
of the type (\ref{slow}). Due to the nonlinear constraint $q(x)^2=1$,
the first term in (\ref{action}) vanishes. The second term can be written,
by undoing the integral over the Dirac fields,
        \begin{equation}
        {\rm e}^{-S[q]} = \int {\cal D}\phi \, {\cal D}\bar\phi \,
        \exp \, - \int d^2x \, \left( \bar\phi^{\rm t} \partial \bar\phi
         + \phi^{\rm t} \bar\partial \phi + \mu \bar\phi^{\rm t} q
        \phi + \mu \phi^{\rm t} q \bar\phi \right) \;.\label{boso}
        \end{equation}  
Here $q$ has to be 
understood as acting like the identity on the spin indices $i=1,\cdots,2N$. 
The free Dirac theory plus the bosonic ghosts is equivalent
to a WZW model with action 
        \begin{equation}
         W_{osp(4N|4N)}[M] = {1\over 16\pi}\int_\CS d^2x\,{\rm STr}\, (M^{-1}
        \partial_\mu M)^2 + {i\Gamma[M] \over 24\pi} \;, \label{wznw}
        \end{equation}   
where the matrix $M$ takes values in a subspace of the complex 
supergroup $OSp(4N|4N)$, and the topological term is expressed
by assuming some extension $\tilde M$ of $M$ to a 3-ball ${\cal B}$
that has position space $\CS$ for its boundary ($\partial{\cal B} =
\CS$)
        \begin{equation}
        \Gamma[M]
         = \int_{\cal B} d^3x \, \epsilon_{\mu\nu\lambda} \, {\rm STr} \,
        \tilde M^{-1}\partial_\mu \tilde M\, \tilde M^{-1}\partial_\nu \tilde M
        \, \tilde M^{-1}\partial_\lambda \tilde M\;.\label{multival}
        \end{equation}
The rules of bosonisation for the
last two terms in the exponent in (\ref{boso}) are to replace
the bilinears $\phi\bar\phi^{\rm t}\Sigma_3$ and $\Sigma_3\bar\phi
\phi^{\rm t}$ by $\ell^{-1} M$ resp.~$\ell^{-1} M^{-1}$, where the factor
$\ell^{-1}$ is a large mass scale, of the
order of $\ell^{-1}_0$, which enters for dimensional reasons.
Up to a conjugation with the matrix ${\rm diag}(1, \Sigma)$, these are 
the same bosonisation rules as in \cite{classD}. 
The term ${\rm STr}\,(M\Sigma_3 q+q\Sigma_3 M^{-1})$
can be viewed as a kind of mass term. At large $\mu/\ell$
it forces the field $M$ to follow $ q \Sigma_3$.
This approximation is valid at momentum scale $k\ll (\mu/\ell)^{1/2}$.
 Neglecting the fluctuations
we can set $M \Sigma_3 q=1$, which yields
        $$
        S[q]\Big|_{q = T\Sigma_3 T^{-1}} =2 N \, W[q\Sigma_3] \;.
        $$    
where the factor $2N$ appears from taking the trace over the 
spin indices. Here $W[q\Sigma_3]$ is the WZW action on $OSp(2|2)$.
Recall that $\Sigma_3\in osp(2|2)$ so that $\pm i\Sigma_3=\exp(\pm i\pi\Sigma_3/2)$
and $q\Sigma_3$ belong to $OSp(2|2)$. 
However, $q\Sigma_3$ does not explore all this group as $q$ is only
a function on the coset space $OSp(2|2)/GL(1|1)$.
Evaluating the topological term for this configuration
can be done by making a smooth extension of $M= q \Sigma_3$
to the ball ${\cal B}$ with radial coordinate $0\leq s \leq 1$,
for example    
         $$
        \tilde M(x,s) = T(x) \exp\left(\pm is\pi\Sigma_3/2\right)
        T(x)^{-1} (\mp i\Sigma_3) \;.
         $$
At $s=1$ we have $\tilde M(x,1) 
 = q(x)\Sigma_3$, while for $s = 0$ we get $\tilde M(x,0)
= \mp i \Sigma_3$, independent of $x$.
Inserting this extension
into the expression (\ref{multival}) for $\Gamma[M]$, and converting
the integral over ${\cal B}$ into an integral over $\CS = \partial
{\cal B}$, we find a theta term
        $$
        {i\over 24\pi} \Gamma[q\Sigma_3] = \pm {1\over 32} \int_\Sigma d^2x \,
        \epsilon_{\mu\nu}\,{\rm STr}\,q\,\partial_\mu q\,\partial_\nu q 
\equiv \pm S_{\rm top}(q)\;.
        $$
Since the value of the WZW topological term does not depend on the 
extension, the two opposite expressions for
$i\Gamma[q\Sigma_3]/24\pi$ are equivalent and
$S_{top}(q)\in i\pi {\bf Z}$.
Gathering the kinetic and topological term, we obtain the following effective 
action
        \begin{equation}
        S[q]\Big|_{q = T\Sigma_3 T^{-1}} =-\frac{2 N}{16\pi}\int d^2x \,
        {\rm STr} \, \partial_\mu q \, \partial_\mu q \\
        \pm  {2N \over 32} \int d^2x \, \epsilon_{\mu\nu} \,
         {\rm STr}\, q \, \partial_\mu q \, \partial_\nu q \;.
        \label{NLsM} 
        \end{equation}
The angle of the theta term is $\theta=\pm 2N \pi$. It
contributes trivially to the path integral as the topological action
is multiplied by $2N$ so that $2NS_{\rm top}\in 2i\pi {\bf Z}$.
The effective action is thus:
\debut
S_{\rm eff}[q]= -\frac{2 N}{16\pi}\int d^2x \,
        {\rm STr} \, \partial_\mu q \, \partial_\mu q \;.
\label{sigmaouf}
\fin
The natural ultraviolet cut-off for this effective action is
$\mu^{-1}\simeq 2\ell_0 e^{1/4g}$ 
since in deriving it we neglected transverse modes
of effective mass $\mu$.

Action (\ref{sigmaouf}) is conjectured to be a massive theory as 
it is a sigma model on a symmetric space with positive curvature.
Recall \cite{Friedan} the one loop renormalization group equations
for sigma model metrics $G_{ab}$:
$$
        l\d_l\;G_{ab} = - R_{ab}
$$
with $R_{ab}$ the Ricci curvature. 
For symmetric spaces the Ricci tensor is proportional to the metric.
We need to compute this proportionality coefficient
in our case. Since $q=T\Sigma_3 T^{-1}$, the tangent space 
at the point $q=\Sigma_3$ is spanned by elements of the form
$[X,\Sigma_3]$ with $X\in osp(2|2)/gl(1|1)$. 
By construction we can choose $X$ such that $\{X,\Sigma_3 \}=0$.
The metric is then:
$$
        G(X,X) = -\frac{1}{8\pi \la} {\rm STr}([X,\Sigma_3])^2
        =\frac{1}{2\pi \la} {\rm STr}(X^2)
$$
where the supertrace is understood in the defining $4$ dimensional
 representation, and $\la=1/2N$.
Similarly, the Ricci tensor at $q=\Sigma_3$ is
defined by \cite{Helga}:
$$
        R(X,X)= - {\rm STr}({\rm ad}\,X)^2
$$
Here the supertrace is in the adjoint representation.
Recall that Ricci tensor are invariant under metric dilatations.
To compute the proportionality coefficient we pick a particular
element of $osp(2|2)$ anticommuting with $\Sigma_3$,
e.g. $X=E_{FF}\otimes \sigma_1$. We have ${\rm STr}(X^2)=-2$.
Diagonalizing the adjoint action, we get a set of eigenvalues zero,
two bosonic non-degenerate eigenvalues $\pm 2$ and
two fermionic eigenvalues $\pm 1$ with multiplicity two.
Hence, ${\rm STr}({\rm ad}\,X)^2=2(2^2-2)=4$. Thus
$G=R/4\pi\la$. The RG equation then becomes:
$$
         l\d_l\,\la = +4\pi \la^2  
$$
At large distance, the model is driven to strong coupling,
and it presumably becomes massive because there is no
contribution from the topological term.

The generated mass scale is of order $m_N\simeq \mu e^{-N/2\pi}\ll \mu$ 
as the coupling constant is equal to $N$ at the ultraviolet cut-off
$\mu$. This is the energy scale at which the coupling constant 
$\la$ becomes of order one.

As a consequence, the infrared fixed point is trivial and 
the zero energy states are localized
with localization length of order $1/m_N$.
In the localized regime, the behavior of the density of states
is expected to be governed by the class $C$ matrix ensemble
\cite{senth}.

\section{Spin-charge separation.}

The conformal field theory with action (\ref{3.3})
admits a spin-charge separation. Its stress tensor
can be decomposed into the sum of the Sugawara stress tensors
associated to the spin and charge current algebras:
\debut
T_{cft} = T_{sp(2N)_0} + T_{osp(2|2)_{-2N}} \label{Tfacto}
\fin
Both Sugawara stress tensors have Virasoro central charge zero
and are bilinear in the corresponding currents:
\debut
T_{sp(2N)_0} &\equiv& \inv{8(N+1)} \lim_{w\to z}\ J^a(z) J^a(w) \non\\
T_{osp(2|2)_{-2N}} &\equiv& -\inv{8(N+1)} \lim_{w\to z} {\rm STr}( K(z)K(w) ) \non
\fin
The normalization of $T_{osp(2|2)_{-2N}}$ may be found in
ref.\cite{SerbMa}. Eq.(\ref{Tfacto}) is proved in Appendix B.

This may be checked by computing the dimension of the 
supermultiplet $\phi$. In the free theory,
its conformal dimension is $1/2$. 
The dimensions in the spin sector are 
$\Delta_{sp(2N)_0} = \frac{{\rm Cas}}{2(N+1)}$ with
Cas. the casimir of the corresponding representation of $sp(2N)$,
cf. ref.~\cite{KZ}. For the vector representation this gives
$\Delta_{sp(2N)_0}=\frac{2N+1}{4(N+1)}$. In the charge sector,
regular representations of $osp(2|2)$ are labeled by two integers
$j,\,b$ and their conformal dimensions are
$\Delta_{osp(2|2)_{-2N}}=\frac{2(j^2-b^2)}{2(N+1)}$, cf. ref.~\cite{SerbMa}.
For the $4$-dimensional representation with $j=1/2,\, b=0$
this gives $\Delta_{osp(2|2)_{-2N}}=\inv{4(N+1)}$.
As it should, the spin and charge conformal dimensions
add up to $1/2$.

For $N=1$, it was shown in ref.\cite{Bhaseen} that the four point
correlation function of the supermultiplet $\phi$ may be factorized
as the product of correlation functions in the $sp(2N)_0$ and
$osp(2|2)_{-2N}$ conformal theories. However, this spin-charge
factorization possesses peculiar properties inherited from 
indecomposability properties of representations of $osp(2|2)$.
In particular, $osp(4N|4N)$ decomposes as:
$$
osp(4N|4N)= osp(2|2)\otimes [{\bf 1}] + [\tilde {\bf 8}]\otimes sp(2N)
+ [{\bf 8}]\otimes [{\bf R}]
$$
with \ $[{\bf R}]$ a $(2N+1)(N-1)$ dimensional representation of
$sp(2N)$ and $[{\bf 8}]$ isomorphic to the adjoint representation of $osp(2|2)$. 
The spin currents $J^a=\phi^t\tau^a\phi$ belong to
$[\tilde {\bf 8}]\otimes sp(2N)$ with $[\tilde {\bf 8}]$ an eight dimensional
indecomposable representation of $osp(2|2)$.
Thus, although these currents commute with the $osp(2|2)$ charge
generators they do not belong to a trivial representation
of the charge algebra.

This separation between spin and charge degree of freedoms still
holds in the perturbed theory provided one fine tunes the coupling
constants such that $g_\al+g_m=g_\al+g_M=0$. In this case:
\debut
 L_{\rm charge} &\equiv& g_\al\, (\CO_\al - \CO_m - \CO_M) \non\\
&=& g_\al\, {\rm STr}[\,
(\,\tau^a\phi \phi^t\tau^a + \inv{2N} \phi \phi^t
+\CT^i\phi \phi^t\CT^i\,)\,\bar \phi \bar \phi^t\,] \non
\fin
where we used again the antisymmetry property (\ref{antisym}).
The sum in the above  r.h.s. projects out the $sp(2N)$ colour indices
leaving only the $osp(2|2)_{-2N}$ currents (\ref{ospcurrent}). Thus
\debut
 L_{\rm charge} = g_\al\, {\rm STr}( K\, \bar K) \label{Lcharge}
\fin
The remaining perturbing operator,
\debut
 L_{\rm spin} \equiv g_A\, \CO_A = g_A\, J^a {\bar J}^a \label{Lspin}
\fin
describes a current interaction involving only the $sp(2N)_0$ generators.
Hence the total perturbation,
\debut
L_{\rm pert} =  L_{\rm spin} + L_{\rm charge} \non
\fin
is then the sum of the two commuting current-current interactions.

The spin-charge separation can also be seen on the beta
functions. When fine tuning coupling constants such that
$g_\al+g_m=g_\al+g_M=0$, the beta functions (\ref{beta1loop}) 
decouple: $\beta_A = \frac{2(N+1)}{N}g_A^2$ 
and $\beta_\alpha = -\beta_m = - \beta_M = -\frac{2}{N}g_\al^2$.
In the fine tuned regime, disorder in the gauge potential $A$ is
marginaly relevant while disorder in the spin potential $\al$ is 
marginaly irrelevant.

\section{Sigma model approach for the spin-charge separated system.}

The line $g_\alpha=g_m=g_M=0$, $g_A =g > 0$ is stable
for the renormalization group flow. It is attractive in the
fine tuned regime $g_\al+g_m=g_\al+g_M=0$, $g_\al>0$, 
and along it the flow is towards
strong coupling, $g_A \to \infty$. This model was
formulated and analysed in \cite{NTW} using replica and
in \cite{beber} by direct means or with supersymmetry. 
Comparison of the various methods was done in \cite{mariage}.
Based on the spin-charge separation, 
it can be deduced \cite{two} that the low energy
physics on this line is given by an $osp(2|2)_{k=-2N}$
theory. Let us derive the same result by using the 
sigma model approach. We shall obtain that the effective
action is a sigma model on the supergroup $OSp(2|2)$, 
endowed with a WZW term, the coupling constants being
such that $k=-2N$. Strictly speaking \cite{classD}, 
the resulting WZW model is defined on a submanifold of the complex
supergroup which is a Riemannian symmetric
superspace of type $D|C$ (meaning that the bosons have an
orthogonal structure and the fermions are symplectic).
Let us remind that by bosonising the free Dirac 
fermions/bosons, one obtains a WZW model on a Riemannian symmetric
superspace of type $C|D$, at level $k=1$.
Since the metric changes sign when passing from a space of type 
$C|D$ to one of type $D|C$, the WZW model on the space of
type $D|C$ is well defined for negative values of the level $k$.  

On the fixed line
$g_\alpha=g_m=g_M=0$, $g_A\neq 0$ the symmetry is not that
of class $C$ any more but that of class $CI$. 
The reason is that the Hamiltonian has now an
extra symmetry 
        $$H=\CT H^T \CT^{-1} \quad {\rm with} \quad 
        \CT=i\sigma_2\otimes \Sigma \;,$$
which can be interpreted as a time reversal symmetry.

The disorder perturbation is now simply the $sp(2N)_0$ 
current-current perturbation. Adding formally terms 
which are zero, we obtain 
        \begin{eqnarray}
        L_{{\rm spin}}&=&\frac{g_A}{2N}\;
        (\bar \phi^t\, \tau^a\, \bar \phi )
        ( \phi^t\, \tau^a\,  \phi ) 
        =\frac{g_A}{2N}\; \str \bar \phi \phi^t 
        E_I (\phi \bar \phi^t )E_I \\
         \nonumber &=&\frac{g_A}{2N}\; \str 
        \big( {\rm Tr}_{gl(2N)} \bar \phi \phi^t \big)
        \big( {\rm Tr}_{gl(2N)} \phi \bar \phi^t )\;,
        \end{eqnarray}
with $E_I$ the (orthonormal) generators of $gl(2N)$.
Since $(\bar \phi \phi^t)^t=-\phi \bar \phi^t$, we can decouple 
this interaction by introducing a supermatrix 
$Q\sim {\rm Tr}_{gl(2N)} \phi \bar \phi^t $. 
We can define an orthosymplectic transposition for $Q$ by 
$Q^t\equiv{\rm Tr}_{gl(2N)}({\bf 1}\otimes Q)^t$.
Remark that in contrast to the preceding section $Q$ has no
specific symmetry properties. In particular 
$Q\neq -Q^t$, so the supermatrix $Q$ belongs to a space larger than 
$osp(2|2)$.  
After decoupling of the interaction term, the effective lagrangian becomes  
        \begin{eqnarray}
        \label{leffCI} 
         L_{\rm eff} &=& \bar\phi^t \partial \bar\phi 
         + \phi^t \bar\partial \phi +2 {\rm STr} \,\( Q 
        {\rm Tr}_{gl(2N)} ( \bar\phi 
        \phi^t)\) + \frac{2N}{g_A} \, {\rm STr} \, Q Q^t  \nonumber \\ 
         &=& (\phi^t\,\bar\phi^t) \pmatrix{\bar\partial &Q\cr 
        -Q^t &\partial\cr} \pmatrix{\phi\cr \bar\phi\cr} - 
        \frac{N}{g_A}\,{\rm STr} \,\pmatrix {0& Q\cr - Q^t&0}^2 \;. 
        \end{eqnarray}
Here, we have embedded $Q$ in the $osp(4|4)$ algebra
represented by matrices of the form
        $$\hat A=\pmatrix{A & B \cr -B^t & D}\;, \quad 
        {\rm with} \quad A^t=-A\;, \quad D^t=-D\;.$$
It is easy to see that the diagonal blocks of $\hat A$ span 
two commuting $osp(2|2)$ algebras, so we conclude that the space to which 
$Q$ belongs is isomorphic to the complement of 
$osp(2|2) \oplus osp(2|2)$ in $osp(4|4)$. 

In the absence of the energy term, the lagrangian (\ref{leffCI}) 
is invariant
under the holomorphic/antiholomorphic transformations 
        \begin{eqnarray}
        \label{lagr}
        \phi \to g_L(z)\;\phi\;&,& \quad \phi^t \to \phi^t\; g_L^{-1}(z)\;,
        \\ \nonumber
        \bar \phi \to g_R(\bar z)\; \bar \phi\;&,& \quad 
        \bar \phi^t \to \bar \phi^t\; g_R^{-1}(\bar z)\;, \\ \nonumber
        Q\to g_L(z)\;Q\; g_R^{-1}(\bar z)\;&,& \quad Q^t\to 
        g_R(\bar z)\;Q^t\; g_L^{-1}(z)\;,
        \end{eqnarray}
with $g_L(z)$ and $g_R(\bar z)$
elements of the $OSp(2|2)$ group: $g_L^t= g_L^{-1}$ and $g_R^t=
g_R^{-1}$. It defines an action of the group
 $G\equiv OSp(2|2)\otimes OSp(2|2)$ on $Q$.
We therefore expect conformal invariance of the $osp(2|2)$ (or charge)
 part of the theory. This is not surprising,
in view of the spin-charge separation, since the charge sector is unperturbed
by the disorder.

Integrating out the Dirac fields, we obtain the effective action 
        \begin{equation}
        \label{effCI}
        S_{\rm eff}=-\frac{N}{g_A}\int \frac{d^2x}{2\pi} 
        \,{\rm STr} \,\pmatrix {0& Q\cr - Q^t&0}^2 + 
        N\; {\bf STr} \ln \pmatrix{ Q\, \Sigma_3& \bar\partial\cr
       \partial&- \Sigma_3\, Q^t\cr}\;.
        \end{equation}
Again, for $N$ large, the integral over the $Q$ field can be treated 
in the saddle point approximation. The diagonal saddle point solution
is also given by a constant matrix of the form $Q= \mu_A \Sigma_3$.
The saddle point equation fixes the value of $\mu_A$,
        $$\mu_A \approx (2\ell_0)^{-1} {\rm e}^{-1/2g_A}\;.$$
Remark that the factor in the exponent has changed compared to
(\ref{genmass}). Due to the fact that, at large $N$, $\beta_A=2g_A^2$, 
$\mu_A$ is also invariant under the RG flow and it has the properties
of a dynamically generated mass.

As usually, the saddle point solution extends to a saddle point
manifold, due to the invariance of the effective action.
This time, the saddle point manifold is generated by
        $$ Q=\mu_A \; g_L(z)\;\Sigma_3\; g_R^{-1}(\bar z)\;,$$
which is a much larger manifold than that of constant matrices.
$Q$ satisfies the non linear constraint $QQ^t=-1$.
Remember that $\Sigma_3^t=-\Sigma_3$.
The stabilizer $H$ of $\Sigma_3$ is constituted of elements
${\rm diag}(g_L,g_R)$
obeying
        $$g_L\;\Sigma_3\; g_R^{-1}=\Sigma_3 \quad \Rightarrow \quad
        g_L=\Sigma_3\; g_R \;\Sigma_3\;.$$
This is compatible with the multiplication law as 
$g_Lh_L=\Sigma_3\; g_Rh_R \;\Sigma_3\;$ for any pairs
$(g_L,g_R)$ and $(h_L,h_R)$. This means that $H\simeq OSp(2|2)$. 
The fluctuations around the saddle point are described
by a sigma model on the coset space $G/H=OSp(2|2)\otimes OSp(2|2)/OSp(2|2)$.
Elements of $G/H$ are pairs $(g_L,g_R)$ with the identification
$(g_L,g_R)\sim (g_L\Sigma_3 h \Sigma_3\;,g_Rh)$ with $h\in OSp(2|2)$.
Convergence of the integrals on the saddle point manifold 
is insured, in the $FF$ sector, by $Q_{FF}^\dagger=-Q_{FF}^t$.
 In the $BB$ sector, convergence requires $g_{L,B}^\dagger=\Sigma_3 
g_{R,B}^{-1}\Sigma_3$, or $Q_{BB}=g_{L,B} g_{L,B}^\dagger\Sigma_3$.
This is equivalent to demand that the
hermiticity condition (\ref{real}) on the bosonic component,
 $\phi^{\dag}_B=\bar \phi^t_B\Sigma_3$, is
preserved by the transformation (\ref{lagr}).
 The even part (or the base) of the saddle point
manifold is then equivalent to the product of two symmetric 
spaces, the fermionic sector being compact and the bosonic one being
non-compact. The full saddle point manifold is a subspace of the
complex group $OSp(2|2)$, being a Riemannian symmetric superspace
of type $D|C$.

In order to perform the gradient expansion, we are making use again
of the non-abelian bosonisation. First, we redefine the 
supermatrix field $Q(x)$
        $$Q(x)=\mu_A\; g(x)\Sigma_3\,,
\quad {\rm with} \quad g=g_L\Sigma_3g_R\Sigma_3 \;.$$
The nonlinear constraint on the new field is $ g^t(x)=g(x)^{-1}$
which is nothing else that the defining relation for an element of the
(complex) supergroup  $OSp(2|2)$. This shows that the coset space
$G/H$ is diffeomorphic to $OSp(2|2)$.
When inserting this expression in the effective lagrangian
(\ref{leffCI}) the last term vanishes. The other terms can be bosonized
using the same rules as in the preceding section. The free Dirac
part gives the WZW action (\ref{wznw}), while the ``mass''
term can be bosonised replacing again $\phi\bar\phi^{\rm t}\Sigma_3$ 
and $\Sigma_3\bar\phi
\phi^{\rm t}$ by $\ell^{-1} M$ resp.~$\ell^{-1} M^{-1}=\ell^{-1}M^t$.
The effective action becomes 
        $$S[Q]=W_{osp(4N|4N)}[M] +\frac{\mu_A}{\ell} 
        \int \frac{d^2x}{2\pi}\str \(g\;M^{-1}+M\;g^{-1}\)\;.$$
The mass term forces $M$ to follow $g$. Neglecting
the fluctuations of $M$ around the minimum $M=g$, we obtain
        \begin{equation}
        \label{conf} 
        S[Q]=2N\;W_{osp(2|2)}[g]\;,
        \end{equation}
where the factor $2N$ appears from taking the trace over the
spin indices, on which $g$ acts as the identity. 
The topological WZW term survives to this reduction
from $OSp(4N|4N)$ to $OSp(2|2)$. The reality conditions discussed
above mean that the restrictions $g_F$ and $g_B$ of $g$ in the FF and BB
sectors satisfy $g_F^{-1}=g_F^\dagger$ and $g_B=g^\dagger_B$,
respectively. This ensures the stability of the action (\ref{conf}).

As already mentioned, $W_{osp(4N|4N)}[M]$ 
is defined on a Riemannian symmetric space of type $C|D$,
while  $W_{osp(2|2)}[g]$ is defined on a Riemannian symmetric 
space of type $D|C$.
The quadratic form defining the metric changes sign between the 
two types of spaces, since $\str\equiv {\rm Tr}_{BB}-{\rm Tr}_{FF}=
{\rm Tr}_{Sp}-{\rm Tr}_{SO}$ in the first case and
$\str=
{\rm Tr}_{SO}-{\rm Tr}_{Sp}$ in the second case. In order to have a
well defined functional integral, the level $k$
of the WZW action has to change sign between the two cases.
We conclude that the action (\ref{conf}) corresponds to an  
$osp(2|2)_{k=-2N}$ theory.
It is surprising that the saddle point approximation is able
to reproduce this exact result.

\section{Appendix A: Algebraic coefficients}
The aim of this appendix is to define the algebraic conventions used
for the calculations. We will also gather various algebraic formulas 
 helpful for reproducing our results for the beta functions.

The ensemble of $sl(2N)$ generators are denoted by $T^A$, and they
can be divided in generators of the subalgebra $sp(2N)$,  $\tau^a$,
and generators of the complement of $sp(2N)$ in $sl(2N)$,  $\CT^i$. 
The normalization we choose is
\debut
Tr(T^A T^B) = \delta^{A B}, & Tr(\CT^i \CT^j)=\delta^{i j},
 & Tr(\tau^a \tau^b)=\delta^{a b} \non
\fin
The various structure constants are defined by
\debut
\left[\CT^i,\CT^j\right] = i{f^{ijk}}\; \CT^k &,& 
\left[\tau^a ,\tau^b \right] = i{f^{a b c}} \;\tau^c \non\\
\left\{\tau^a ,\tau^b \right\} = i{b^{a b i}}\; \CT^i\;+\; x^{a b} I&,&
\left[\CT^i,\CT^j\right] = i{f^{ija}}\; \tau^a \non\\
\left[\tau^a ,\CT^i\right] = i{f^{a i j}}\; \CT^j &,&
\left\{\tau^a ,\CT^i\right\} = i{b^{a i b}}\; \tau^b \non\\
\left[T^A ,T^B \right] &=& i{f^{A B C}}\; T^C\non
\fin
where $[.,.]$ stands for commutators and $\{.,.\}$ anticommutators.

When evaluating beta-functions, a certain number of algebraic
identities are needed to simplify expressions.
Here is a list of algebraic coefficients that are needed :

\begin{center}
\begin{tabular}{|c|c|c|}
\hline
Coefficient & Identity & Value \\
\hline
\hline
${\cal{C}}$ & $T^a T^A = {\cal{C}} $ & $(4N^2-1)/2N$ \\
\hline
$C$ & $\tau^a \tau^a = C $ & $(2N+1)/2$ \\
\hline
$C'$ & $\CT^i \CT^i = C'$ & $(2N+1)(N-1)/2N$ \\
\hline
$x_{TTT}$ & $ T^A T^B T^A = x_{TTT} T^B$ & $-1/2N$ \\
\hline
$x_{\tau\tau\tau}$ & $\tau^a \tau^b \tau^a = x_{\tau\tau\tau}\tau^b$ & $-1/2$ \\
\hline
$x_{\CT\tau\CT}$ & $\CT^i\tau^a\CT^i = x_{\CT\tau\CT}\tau^a $ & $(N-1)/2N$ \\
\hline
$x_{\tau\CT\tau}$ & $\tau^a\CT^i\tau^a = x_{\tau\CT\tau}\CT^i$ & $1/2$ \\
\hline
$x_{\CT\CT\CT}$ & $\CT^i\CT^j\CT^i=x_{\CT\CT\CT}\CT^j$ & $-(N+1)/2N$ \\
\hline
$D_{TT}$& $f^{CDA}f^{CDB}=-D_{TT}\delta^{AB}$ & $4N$\\
\hline
$D_{\tau\tau}$ & $f^{cda}f^{cdb}=-D_{\tau\tau}\delta^{ab}$ & $2(N+1)$\\
\hline
$D_{\CT\CT}$ & $f^{ija}f^{ijb}=-D_{\CT\CT}\delta^{ab}$ & $2(N-1)$ \\
\hline
$D_{\tau\CT}$& $f^{aki}f^{akj}=-D_{\tau\CT}\delta^{ij}$ & $2N$ \\
\hline
$B_{\tau \tau}$& $b^{abi}b^{abj}=-B_{\tau \tau}\delta^{ij}$ & $-2(N+1)$ \\
\hline
$B_{\tau\CT}$ & $b^{cia}b^{cib}=-B_{\tau\CT}\delta^{ab}$ & $-2(N^2-1)/N$ \\
\hline
$x$ & $x^{a b}x^{a b} = x$ & $(2N+1)/N$\\
\hline
\end{tabular}
\end{center}

\section{Appendix B: Stress tensor factorization.}
The decomposition (\ref{Tfacto}) may be verified directly by
computing the operator products defining the stress tensors using 
Wick theorem with the normalization (\ref{wick}). For the
$sp(2N)$ sector we get:
\debut
T_{sp(2N)_0}&=& \inv{(N+1)} \lim_{w\to z}\ [ 
\inv{2(z-w)}(\phi^t_{(z)}\tau^a\tau^a\phi_{(w)})
+\inv{8} (\phi^t\tau^a\phi)^2_{(z)}\, ] \non\\
&=& -\frac{2N+1}{4(N+1)} (\phi^t\d\phi) + \inv{8(N+1)}
(\phi^t\tau^a\phi)^2 \non
\fin
where we used $\phi^t\phi=0$, by antisymmetry, and 
the value of the $sp(2N)$ Casimir in the defining
representation, $\tau^a\tau^a=(2N+1))/2$.
Similarly for the $osp(2|2)$ sector we first introduce the appropriate
basis of $gl(2N)$ to decompose $T_{osp(2|2)_{-2N}}$ as follows:
\debut
 T_{osp(2|2)_{-2N}}= &-& \inv{8(N+1)}\lim_{w\to z} 
{\rm STr}\left({(\phi\phi^t)_{(z)}\, E_I \,
(\phi\phi^t)_{(w)} E_I }\right)\non\\
=&-&\inv{8(N+1)}\lim_{w\to z}\ [
 (\phi^t_{(z)}\tau^a\phi_{(w)})(\phi^t_{(w)}\tau^a\phi_{(z)}) \non\\
&+&\inv{2N}(\phi^t_{(z)}\phi_{(w)})(\phi^t_{(w)}\phi_{(z)})
+(\phi^t_{(z)}\CT^i\phi_{(w)})(\phi^t_{(w)}\CT^i\phi_{(z)}) \,]\non
\fin
Because of the antisymmetry property,
$\phi^t\phi=\phi^t\CT^i\phi=0$, eq. (\ref{antisym}),
the two last terms are proportional to $(\phi^t\d\phi)$.
The first term gives a contribution 
proportional to $(\phi^t\d\phi)$ but
also to $(\phi^t\tau^a\phi)^2$. Gathering these contributions
and using again the value of the $sp(2N)$ Casimir as well as
$\CT^i\CT^i=(2N+1)(N-1)/2N$, we get:
\debut
T_{osp(2|2)_{-2N}} =  -\frac{1}{4(N+1)} (\phi^t\d\phi) 
-\inv{8(N+1)} (\phi^t\tau^a\phi)^2  \non
\fin
Hence:
\debut
T_{sp(2N)_0}+T_{osp(2|2)_{-2N}}= -\inv{2}  (\phi^t\d\phi) 
\equiv T_{cft} \non
\fin
which proves the claimed factorization.

\section{Appendix C: All order beta functions.}

In this appendix we derive expressions for the beta functions using
formula suggested in ref. \cite{Aetal}. These formula apply
to current-current perturbations of WZW models of the form:
$$
S= S_{\rm wzw} + \sum_K h_K \int\frac{dx^2}{2\pi} \CO^K
\quad,\quad
\CO^K= d^K_{a\bar a} J^a J^{\bar a}
$$
where $J^a$ and $J^{\bar a}$ are the left and right conserved
currents of the WZW models generating an affine (super) algebra 
at some level $k$. In the case of a superalgebra the currents $J^a$
possess a bosonic or fermionic character depending whether their
degree $|a|$ are zero or one.
In our case, the underlying algebra is  the affine $osp(4N|4N)$
at level $k=1$. The currents are bilinear in the supermultiplet
$J^a = \phi^t X^a \phi$,  with $X^a$ generators of $osp(4N|4N)$.

For the theory to be  perturbatively renormalizable,
one needs to choose the tensors
$d^K_{a\bar a}$ such that:
\debut
(-)^{|b||c|} d^K_{ab}d^L_{cd} f^{ac}_if^{bd}_j &=& C^{KL}_I d^I_{ij}\non\\
\eta^{ij}d^K_{ai}d^L_{bj} &=& D^{KL}_I d^I_{ba} \non\\
d^K_{ij} f^{ja}_kf^{ik}_b &=& R^K_L \eta^{ac}d^L_{cb} \non
\fin
where $\eta^{ab}$ is the Killing invariant bilinear form of 
the superalgebra and $f^{ab}_c$ its structure constants.
These conditions are satisfied by the four operators
$\CO_\al,\ \CO_A,\ \CO_m$ and $\CO_M$ defined in eq. (\ref{oppert}).

With an appropriate renormalization prescription,
the proposed beta functions \cite{Aetal} read for $k=1$,
\debut
\beta_h = - C(h',h')(1+D(h)^2) +2C(h'D(h),h'D(h)) -2 h'D(h)RD(h)
\label{allbeta}
\fin
where $D(h)$ is the matrix $D(h)^K_L=D^{KI}_L h_I$,
$C(x,y)$ is the row vector with components $C(x,y)_K=C^{LI}_K x_Ly_I$,
and finally
$$
h' = h (1 - D(h)^2 )^{-1}
$$
We shall assume that these formula are correct and capture 
perturbative contributions to all orders.

With the normalization of eq.(\ref{3.2}), $h_K=g_K/2N$.

One has to compute all the tensors $C^{KL}_I$, $D^{KL}_I$ and $R^K_L$.
The first one codes the operator product expansion of the
perturbing operators:
\debut
\CO^K(z)\CO^L(0) \simeq \inv{|z|^2} \sum_C C^{KL}_I \CO^I(0) 
+ {\rm reg} \non
\fin
The coefficients $C^{KL}_I$ have been determined when computing the 
one-loop beta functions, see eq.(\ref{beta1loop}).

The tensor $D^{KL}_I$ may be computed using a variation on the free field
representation  of the currents. Namely introduce
auxiliary copies $\phi_\al$, $\al =0,1\ {\rm or}\ 2$
 of the supermultiplets with two point
function $\vev{\phi_\al\phi^t_\al}_\al={\bf 1}$. 
Here $\vev{\cdots}_\al$ denotes the expectation value in the auxiliary
Fock space associated the the auxiliary field $\phi_\al$. Let 
$$
\CO^K_{\al\beta} \equiv d^K_{a \bar a} (\phi_\al^tX^a\phi_\al)
(\phi_\beta^tX^{\bar a}\phi_\beta)
$$
Then one has:
$$
\vev{\CO^K_{01} \CO^L_{02}}_0 = D^{KL}_I \CO^I_{12}
$$
Hence, $D^{KL}_I$ is computable simply using Wick theorem.

The last tensor $R^K_L$ is computable by looking at the
following operator product expansion:
\debut
T^K(z)\CO^L(0)\simeq \frac{1}{z^2} 
(2D^{KL}_I + R^K_N D^{LN}_I)\CO^I(0) + {\rm reg}  \non
\fin
with $T^K(z)=d^K_{ab} J^a(z)J^b(z)$.\\ 
The result is summarized in the following tables:
\debut
&& D^{mm}_m=D^{Am}_m =D^{\al m}_\al = D^{Mm}_M =1/N,
\quad D^{MM}_m =(2N+1)(N-1)/N \non \\
&& D^{\al A}_A =1,\quad D^{\al \al}_M=-D^{\al \al}_\al=N+1,
 \quad D^{MM}_M=(N+1)(N-2)/N \non\\
&& D^{\al \al}_m=2N+1,\quad D^{MM}_\al=-N+1, \non\\
&& \quad D^{M A}_A= (N-1)/N,\quad D^{M\al}_M=-N,
 \quad  D^{M \al}_\al = (N^2-1)/N \non
\fin
The non-vanishing $F^{KL}_I=2D^{KL}_I +R^K_N D^{LN}_I$ are:
\debut
&& F^{AA}_A=2F^{A\al}_M=2F^{AM}_M=2F^{\al\al}_M=8(N+1)\non\\
&&F^{Am}_m=F^{A\al}_m=F^{\al\al}_m=-2F^{\al m}_m=4(2N+1) \non\\
&& F^{Am}_A=F^{Am}_\al/2=F^{\al m}_A=F^{\al m}_\al/2=
-F^{mA}_A=-F^{mm}_m=-F^{m\al}_\al=-F^{2M}_M=2/N \non\\
&& F^{A\al}_A=F^{\al\al}_A=2,\quad
F^{MA}_A=F^{Mm}_m=F^{M\al}_\al=F^{MM}_M= -2(2N+1)(N-1)/N \non\\
&&F^{A\al}_\al=4N,\quad F^{AM}_A=F^{\al M}_A=2(N-1)/N,
\quad F^{AM}_\al=F^{\al M}_\al=4(N^2-1)/N \non\\
&& F^{\al A}_A=-2(2N-1),\quad
F^{\al\al}_\al =-2(4N+3),\quad F^{\al M}_M=-2(4N+1) \non
\fin
For arbitrary $N$, the beta functoins are then derived 
from eq.(\ref{allbeta}). In the $N=\infty$ limit, they  reduce to:
\debut
\beta_M&=&\beta_\al= 2(g_\al+g_M)g_A \label{Betc}\\
\beta_A &=& 2g_A^2 + (g_\al+g_M)^2/2\non\\
\beta_m&=& 4g_A(g_\al+g_m)-2g_m(g_\al+g_M) 
- \frac{(g_\al+g_M)(g_\al-g_M)^2}{(g_\al-g_M-2)}  \non
\fin
These are easy to integrate. In particular $(g_\al-g_M)$
is a RG invariant in the large $N$ limit while the beta functions
for 2$g_A\pm(g_\al+g_M)$ are separated and quadratic.

It is then simple to verify that the isotropic line
$g_\al=g_m=g_M=g_A\equiv g$ is stable and attractive to all orders
and $\beta_g= 4 g^2$. The fact that it is quadratic is in
agreement with eq.(\ref{gsaddle}) and it provides a tiny check
of the all order beta functions. The isotropic coupling grows
indefinitely with the scale. Of course the large $N$ approximation remains
valid only for $g\ll N$. It is possible to verify that
all RG trajectories starting in the domain of positive couplings
sufficiently close to the origin are asymptotic to the isotropic
line at large distances. This confirms the one loop computation.

\end{document}